\documentclass[superscriptaddress,nofootinbib,amsmath,amssymb,aps,pra,twocolumn,article,longbibliography,floatfix]{revtex4-2}

\usepackage{graphicx}
\usepackage{dcolumn}
\usepackage{bm}
\usepackage{hyperref}
\usepackage{braket}
\usepackage[mathlines]{lineno}

\hypersetup{breaklinks=true,colorlinks,linkcolor=black,citecolor=blue,urlcolor=blue}

\begin{document}

\title{How to avoid the appearance of a classical world in gravity experiments}
\thanks{This paper is dedicated to Dieter Zeh and his remarkable work.}
\author{Markus Aspelmeyer}
\affiliation{Faculty of Physics, University of Vienna, Boltzmanngasse 5, 1090 Vienna, Austria}
\affiliation{Institute for Quantum Optics and Quantum Information (IQOQI) Vienna, Austrian Academy of Sciences, Boltzmanngasse 3, 1090 Vienna, Austria}
\email{markus.aspelmeyer@univie.ac.at}


\begin{abstract}
Quantum states of gravitational source masses can lead to experimental outcomes that are inconsistent with the predictions of a purely classical field theory of gravity. Environmental decoherence places strict boundary conditions to the potential realization of such experiments: sufficiently mild not to act as a fundamental show-stopper, yet sufficiently demanding to represent a formidable challenge to the next generation of quantum experiment(er)s.
\end{abstract}

\maketitle

\section{Introduction}
There is renewed and growing interest in producing hard experimental evidence that gravity requires a quantum description. The origins of this quest can be traced back to at least the Chapel Hill Conference in 1957 on ``The role of gravitation in physics". At this conference, to make the point \textit{``that we're in trouble if we believe in quantum mechanics but \emph{don't} quantize gravitational theory"} \cite{DeWitt2011}, Feynman proposed a gedankenexperiment with macroscopic superpositions of massive objects that are coupled via gravity. Since the interaction involves superpositions of the gravitational field, a correct description of the experiment would also have to involve a quantum description of gravity. More than 50 years later, Claus Kiefer and I were part of a working group on 
``Historical Roots of Quantum Gravity Research" at the Max Planck Institute for the History of Sciences, where the newly published transcript of the Chapel Hill conference was heavily discussed during several sessions in Berlin \cite{Kiefer2013}. Claus brought the transcript to the attention of Dieter Zeh, who responded with a fantastic comment on Feynman's contributions \cite{Zeh2011}. I am following up on this debate by addressing a question that was asked by another Chapel Hill participant, Louis Witten: \textit{
``What prevents this from becoming a practical experiment?"}. 

Feynman's initial response to Witten's question was clearly in the spirit of Zeh's decoherence: \textit{``You might argue this way: Somewhere in your apparatus this idea of \emph{[probability]} amplitude has been lost. You don’t need it any more, so you drop it. The wave packet would be reduced (or something). Even though you don’t know where it’s reduced, it’s reduced. And then you can’t do an experiment which distinguishes interfering alternatives from just plain odds (like with dice)."} In other words: the attempt to amplify quantum coherence to the level of macroscopic superpositions will likely result in loss of information to the environment. It is fair to say that at that time there was no prospect for actual experiments that involve gravitational source masses in the quantum regime. 
Amazingly, this has changed. Looking at today's landscape of quantum experiments involving objects of increasing mass and complexity in spatial superpositions \cite{Hornberger2012,Arndt2014} and the fast progress in quantum controlling the motion of solid state objects \cite{Aspelmeyer2014} such experiments seem to be possible in principle - provided that we manage to prevent decoherence from happening too fast.

\section{Feynman's gedankenexperiment}
The original idea put forward by Feynman is simple: \textit{``One should think about designing an experiment which uses a gravitational link and at the same time shows quantum interference"} \cite{DeWitt2011}. Concretely, a Stern-Gerlach magnet produces a spatial superposition of a gravitational source mass carrying a net spin of 1/2. The source mass state is then described by $\ket{\Psi_0}_s=\ket{x_L}_s+\ket{x_R}_s$, with $x_{L,R}$ being two positions (say left, right) of the center of mass along the x-axis that are macroscopically distinct ($\braket{x_L|x_R}_s\ll 1$). In a next step, the delocalized mass couples via gravity to a distant test mass $\ket{\Phi_0}_t=\ket{x_0}_t$ at location $x_0$. If quantum mechanics holds, the test mass experiences two possible accelerations $a_{L,R}=G m_s/(|x_{L,R}-x_0|^2)$ ($G$: Newton's constant; $m_s$: source mass) that result in entanglement between the masses once the separation in test mass displacement is sufficiently large, i.e. $\ket{\Psi_0}_s\otimes\ket{\Phi_0}_t\rightarrow\ket{x_L}_s\ket{x_L}_t+\ket{x_R}_s\ket{x_R}_t$ for $\braket{x_L|x_R}_t\ll 1$. 

From a non-relativistic quantum physics point of view the generation of entanglement is a trivial consequence of gravitational coupling between the two masses, where each branch of the superposition is treated separately. From a field point of view, the generation of entanglement requires the gravitational field of the source mass configuration to be in a quantum superposition. This, however, poses a problem for general relativity. Such a superposed field configuration is not a valid solution of Einstein’s field equations of gravity\cite{Unruh1984}. Simply said: a classical field theory cannot produce a quantum superposition of fields because it cannot deal with sources that are placed in superpositions to start with. This is the case discussed by Feynman.\footnote{Feynman uses a somewhat more drastic wording: 
``\textit{But aside from that possibility} [that quantum mechanics fails]\textit{, if you believe in quantum mechanics up to any level then you have to believe in gravitational quantization in order to describe this experiment.}". Here he explicitly refers to the ability to apply the superposition principle, i.e. probability amplitudes, to different gravitational field configurations, and \emph{not} the necessity to invoke quanta of the field (gravitons).} 
This incompatibility also emerges on the level of an effective Hamiltonian description: for instantaneous interactions, entanglement is generated by the classical Newtonian potential. If one takes into account that the interaction is mediated locally via the gravitational field, entanglement between the masses can only be created when the mediator itself is genuinely non-classical (see e.g. \cite{Bose2017,Marletto2017,Marshman2020,Marletto2020}, and \cite{Galley2020} for a more general no-go theorem). It is also instructive to analyze the situation from the geometric viewpoint of general relativity, in which gravitational interaction is understood as geodesic motion in a non-Minkowski space-time metric. In the low-energy limit of solar system dynamics or table-top experiments this is effectively described by $g_{00}$ being sourced by the mass density configuration $\varrho\left(\underline{r}\right)$. The Newtonian potential $\phi$ is then simply given by the flat-space perturbation $h_{00}=1-g_{00}$ as $\phi=-\frac{1}{2}h_{00}=-G\frac{M}{r}$. A test mass freely falling in a fixed metric will not be able to become entangled in either position or momentum. Entanglement can only occur if the space-time metric itself is in a quantum superposition \cite{Christodoulou2019}.  

So by saying that the generation of entanglement in Feynman's gedankenexperiment requires a quantum description of the gravitational field, what one actually means is that the observation of entanglement is inconsistent with the assumption of a purely classical source mass configuration -- and hence with a well-defined space-time metric. At the same time, the outcome will likely be fully consistent with the quantum theoretical predictions based on an effective quantum field theory in the linearized regime \cite{Donoghue1994,Blencowe2013,Belenchia2018}. This is reminiscent of quantum optics experiments in the early 1970s that were designed to exclude semiclassical theories of radiation \cite{Clauser1972,Clauser1974}, and that culminated in the seminal Bell experiments \cite{Freedman1972,Aspect1982,Weihs1998} to rule out the broad class of local-realistic models as an underlying description for quantum phenomena \cite{Bell1964}. Violating a Cauchy-Schwartz inequality in a photon correlation experiment does not imply quantization of the electromagnetic field; it rather excludes the possibility of being describable by a genuinely classical theory, while at the same time being consistent with the quantum field theoretic predictions of electrodynamics \cite{Clauser1974}. Similarly, gravity experiments whose outcome cannot be explained by a purely classical source mass configuration would rule out the possibility of genuinely classical theories of gravity \cite{Page1981,Unruh1984}. One specific example is semiclassical gravity, in which the quantum mechanical energy-momentum operator $\hat{T}_{\mu\nu}$ is replaced by its expectation value $\braket{\hat{T}_{\mu\nu}}$ as a source for the gravitational field \cite{Moller1962,Rosenfeld1963}. As a consequence, the space-time metric is always well defined -- since even a mass in a superposition would contribute an actual energy density at each of its possible locations in the superposition -- and can hence never reproduce phenomena due to quantum interference\footnote{Note that this is also one underlying assumption in gravitational collapse hypotheses as formulated by Penrose \cite{Penrose2014} and others \cite{Diosi1984,Colin2014}. Therefore any observation of gravitationally induced entanglement would rule out the validity of these models by principle. Some quantum theories of gravity also expect a breakdown of the superposition principle\cite{Stamp2015,Barvinsky2021}, but they have not yet been analyzed in the context of the discussed experiments.}. Formally, the resulting dynamics is captured by the non-linear Schr\"odinger-Newton equation \cite{Bahrami2014}, whose predictions for massive quantum objects provide an independent testbed for semiclassical gravity \cite{Giulini2011,Yang2013,Colin2016}.

In the following we will focus on the actual challenge when trying to implement such experiments: to maintain full control over quantum coherence at the level of the gravitational source (and test) masses. This can only be achieved by sufficiently isolating the experiment from environmental influences such that all decoherence rates are small compared to the time scales of the wanted gravitational coupling. Otherwise, environmental decoherence \cite{Joos1985} will result in a mass distribution that cannot, even in principle, be distinguished from an incoherent mixture and therefore will always be consistent with semiclassical gravity (as is for example the case in the first experimental attempt of Page and Geilker \cite{Page1981}).

\section{Witten's challenge: becoming a practical experiment}
We restrict the discussion to the generation of quantum entanglement via gravitational interaction, as it is at the heart of most previous debates \cite{DeWitt2011,Page1981,Unruh1984}. Two concrete possibilities have been recently proposed in the literature: (i) Entanglement via gravitational phase evolution of superposition states \cite{Bose2017,Marletto2017}, which is an elegant adaptation of Feynman's thought experiment, and (ii) entanglement via gravitational interaction of two quantum harmonic oscillators \cite{AlBalushi2018,Krisnanda2020,Weiss2020}. Since both cases deal with table-top experimental settings, it is sufficient to use the linearized, low-energy regime of general relativity, i.e. Newtonian dynamics. For both approaches we consider two masses $m$ that are kept at a fixed distance $d$ along the z-axis and that are free to move along one orthogonal direction (x-axis). The particles are prepared in independent initial states $\ket{\Psi_0}_1, \ket{\Phi_0}_2$ and are assumed to interact via gravity only.

In the first scenario \cite{Bose2017,Marletto2017} we start out with each of the masses being prepared in a spatial superposition state of size $\Delta x=|x_L-x_R|$, i.e. $\ket{\Psi_0}_1=\ket{x_L}_1+\ket{x_R}_1$ and $\ket{\Phi_0}_2=\ket{x_L}_2+\ket{x_R}_2$. Again, $x_{L,R}$ are two macroscopically distinct positions of the center of mass of each particle along the x-axis, i.e. $\braket{x_L|x_R}\ll 1$. In the presence of a gravitational potential $\phi$ each particle wavefunction picks up a dynamical phase $\varphi=\frac{1}{\hbar}\int m\phi\,dt$. This was used in the seminal 
``COW" experiment by Colella, Overhauser and Werner to generate quantum interference of neutrons induced by Earth's gravitational field \cite{Colella1975,Overstreet2022}. Here, in contrast, the source for the gravitational field is one of the particles. The potential experienced by the other particle can take the values $\phi_{0,1}=Gm/|r_{0,1}|$ ($r$: center-of-mass distance) with  $|r_0|=d$ and $|r_1|=\sqrt{d^2+\Delta x^2}$. Since the source mass is in a superposition state, the resulting conditional 
``COW" phase shifts $\varphi_{0,1}(t)=\frac{1}{\hbar}m\phi_{0,1}t$ can generate entanglement between the previously independent particles\footnote{For $\Delta\phi=\pi$ this is known in quantum information as an entangling CSIGN gate. \cite{Nielsen2010}}. Specifically, the evolution of the 2-particle state is given by $\ket{\Psi_0}_1\otimes\ket{\Phi_0}_2\rightarrow\ket{x_L}_1\ket{\tilde{x}_L}_2+\ket{x_R}_1\ket{\tilde{x}_R}_2$, with $\ket{\tilde{x}_L}_2=\ket{x_L}_2+e^{i\Delta\varphi}\ket{x_R}_2$,  $\ket{\tilde{x}_R}_2=e^{i\Delta\varphi}\ket{x_L}_2+\ket{x_R}_2$ and $\braket{\tilde{x}_L|\tilde{x}_R}_2\propto |1+\mathrm{cos}(2\Delta\varphi)|$, where $\Delta\varphi=\varphi_1-\varphi_0$. As a result, entanglement is generated for $\Delta\varphi>0$ and at a rate $\Gamma_{\mathrm{ent}}=\frac{d}{dt}\Delta\varphi\approx(G/\hbar)(m^2\Delta x^2/d^3)$, consistent with \cite{Bose2017,Marletto2017}.

In the second scenario, two harmonically bound masses are prepared in pure quantum states with wavepacket extensions $\Delta x=\eta\sigma_0$ ($\eta$: expansion parameter; $\sigma_0=\sqrt{\hbar/(m\omega_0)}$: quantum ground state position uncertainty; $\omega_0$: trap frequency). In most cases, this will likely be realized by squeezed states of motion 
quantified by $\eta=e^{s}$, where $s$ is the corresponding squeezing parameter. 
Newtonian gravity facilitates a $1/r$ coupling between the masses via the Hamiltonian $\hat{H}_G=-Gm^2/|\hat{r}|$, with the center of mass distance $|\hat{r}|=\sqrt{d^2+|\hat{x}_1-\hat{x}_2|^2}$. Taylor expansion yields the first relevant entangling interaction term $\hat{H}_{\mathrm{int}}\approx(Gm^2/d^3)\hat{x}_1\hat{x}_2=\hbar\Gamma_{\mathrm{ent}}\hat{x}_1\hat{x}_2/\Delta x^2$, with $\Gamma_{\mathrm{ent}}\approx(G/\hbar)(m^2\Delta x^2/d^3)$ being the rate at which Gaussian entanglement is created\footnote{Higher-order terms of the expansion likely yield non-Gaussian contributions to the entanglement formation.} and consistent with \cite{AlBalushi2018,Krisnanda2020,Weiss2020}.

Somewhat surprisingly, even though both scenarios are experimentally very different they generate entanglement at the same rate $\Gamma_{ent}$  (in the ideal case of pure states). The relevant parameters are the mass $m$, the center-of-mass distance $d$ and the size of the delocalization $\Delta x$ of each of the masses. For scenario 1, $\Delta x$ is the separation between two center-of-mass states of a spatial superposition, for scenario 2 it is the spatial extent of a Gaussian wavepacket. To provide an example: if we assume that we can keep two masses at a distance $d\approx 1\mu$m and can prepare a delocalization $\Delta x\approx 100$nm, it would require at least $N\approx10^{11}$ silicon atoms to achieve entanglement rates on the order of seconds. Given that the confinement of these atoms needs to be on a scale much smaller than $d^3$ this will require solid-state densities for each mass. In the following we will therefore restrict the discussion to quantum states of motion of solids (rather than atomic gases or molecules) and to the idealized case of spherical masses. In order to avoid unphysical constraints (center-of-mass distance smaller than object sizes) we parametrize the center-of-mass distance via $d=2R\alpha$, with particle radius $R$ and $\alpha>1$ (e.g. $\alpha=2$ corresponds to a surface-to-surface distance between the masses of $2R$). Using $m\approx 4R^3\rho$ ($\rho$: material density) this results in $\Gamma_{\mathrm{ent}}\approx (G/\hbar)(2\rho^2/\alpha^3)R^3\Delta x^2$. Coming back to Witten's question: it is obvious that environmental decoherence will be able to prevent this from becoming a practical experiment. So what are the sources of decoherence and how can we overcome them?

\section{Zeh's decoherence: avoiding the appearance of a classical world}
Let us start again with scenario 1, the ``CSIGN" approach to gravitational entanglement. As of today it is unclear how well one can realize the initial state $\ket{\Psi_0}=\ket{x_L}+\ket{x_R}$ that is required for this protocol, namely a superposition of macroscopically distinct states of the center of mass position of a massive object. The trend in existing experiments indicates a clear trade-off between separation $\Delta x$ and mass $m$. For example, while experiments with individual neutrons or atoms achieve separations over tens of centimeters \cite{Colella1975,Kovachy2015}, experiments with macromolecules comprising up to $2,000$ atoms achieve separations over hundreds of nanometers \cite{Fein2019}, and experiments with solid-state resonators made of more than $10^{12}$ atoms separations only on the level of femtometers \cite{O'Connell2010}\footnote{To state a fun fact: in all of the existing macroscopic quantum experiments the trade-off between mass and superposition size is of the same order of magnitude with $\Delta x\cdot m \approx 10^{-28}$ kg$\cdot$m.}. There are proposals in the literature to overthrow this trend, both for dielectric \cite{Romero-Isart2011,Bateman2014} and superconducting \cite{Romero-Isart2012,Pino2016} solids, however thus far no experiment has entered these extreme regimes of macroscopic quantum superpositions. For now we assume that the preparation of such states is possible and ask for the minimum requirements on the initial state to implement the entanglement protocol. A first strict bound is provided by the achievable vacuum level in the experiment. Cryogenic XUV experiments operating at environment temperatures $T_e\approx 1$K can achieve gas densities that correspond to pressure levels below $10^{-17}$mbar \cite{Gabrielse2001}. Typical de-Broglie wavelengths of hydrogen molecules, the dominating rest gas species in this regime, are on the order of $\lambda_{th}=2\pi\hbar/(\sqrt{2\pi m_{H2}k_BT_e})\approx 1$nm, which is likely much smaller than the superposition size $\Delta x$. This means that scattering of a single molecule is sufficient to localize the superposition state. As a consequence, decoherence from rest gas scattering occurs at a rate \cite{Joos1985,Romero-Isart2011a} $\Gamma_g=(\lambda_{th}/\hbar)(16\pi/3)pR^2\approx 2\times 10^{26}pR^2$ ($p$: gas pressure, $R$: particle radius). To generate entanglement in the presence of rest gas requires $\Gamma_{\mathrm{ent}}>\Gamma_g$, which reduces to the condition $R\Delta x^2>10^{26}(\hbar/G)(\alpha^3/\rho^2)p\approx 10^{-12}(1/\rho^2)$ for $p=10^{-17}$mbar=$10^{-15}$Pa and $\alpha=2$. Ideally, one chooses large particles at large delocalization, since the mass coupling ($\propto R^3$) grows faster than the decoherence due to scattering  ($\propto R^2$). Practically, as is mentioned above, there will be a trade-off between those two numbers. Looking at current experiments, silica nanoparticles ($\rho\approx 2\times10^3$ kg$\cdot$m$^{-3}$) with $R\approx75$nm have now been prepared in their quantum ground state of motion\cite{Delic2020,Magrini2021,Tebbenjohanns2021}. They could provide a promising starting point for this experiment, albeit at a required superposition size of $\Delta x>2\mu$m, which seems challenging. At the other end of the spectrum, a Planck-mass lead particle ($R\approx 70\mu$m, $\rho\approx 10^4$ kg$\cdot$m$^{-3}$) would require $\Delta x>12$nm, but it is unclear as of yet if quantum control can be extended to this mass regime. \\
Another omnipresent source of decoherence is scattering, absorption and emission of blackbody radiation, which provides a temperature limit to both the environment temperature $T_e$ (for scattering and absorption) and the internal particle temperature $T_i$ (for emission). The thermal wavelength of the photons $\lambda_{th}^{e,i}=\pi^{2/3}\hbar c/(k_B T_{e,i})\approx 5$mm for $T_{e,i}\approx 1$K is significantly larger than $\Delta x$ and hence decoherence occurs at a rate $\Gamma_{bb}=\Lambda_{bb}\Delta x^2$. Scattering of blackbody photons is described by \cite{Joos1985,Romero-Isart2011a} $\Lambda_{bb}^{sc}=(1/\lambda_{th}^e)^9(8!8\zeta(9)\pi^5cR^6/9)\mathrm{Re}\{(\epsilon-1)/(\epsilon+2)\}^2\approx 5\times 10^{36}R^6T_e^9$ ($\zeta$: Riemann Zeta function, $\epsilon$: dielectric constant), and emission and absorption by $\Lambda_{bb}^{e,i}=(1/\lambda_{th}^{e,i})^6(16\pi^9cR^3/189)\mathrm{Im}\{(\epsilon-1)/(\epsilon+2)\}\approx 5\times10^{25}R^3T_{e,i}^6$. Here we have assumed the worst case scenario of large dispersion and absorption, for which $(\epsilon-1)/(\epsilon+2)\approx 1$. Using again the examples from above, for a $75$nm silica particle $\Gamma_{\mathrm{ent}}>\Gamma_{bb}$ results in bounds $T_{i,e} < \left[(G/\hbar)(2\rho^2/\alpha^3)(1.4\times10^{-26})\right]^{1/6} \approx 5$K from blackbody emission and absorption, and $T_e < \left[(G/\hbar)(2\rho^2/\alpha^3)(1/R^3)(3\times10^{-37})\right]^{1/9}\approx 40$K  from blackbody scattering. For a Planck mass lead particle, one obtains $T_{i,e}<8$K from emission and absorption, and $T_e<6$K from scattering. The limits on internal temperature are independent of the object size and ultimately set by the material absorption properties. In the case of levitated superconductors or magnets, small internal temperatures should be easily achievable as state preparation will likely occur at cryogenic temperatures in a magnetic trap \cite{Pino2016,Gieseler2020,Latorre2021}, while optically levitated particles will need to implement either active cooling methods or switch to 
``dark" traps such as  electrostatic traps \cite{Conangla2018,Delord2020,Dania2021}. On the other hand, the available environment temperature will set an ultimate bound on the object size. For the extreme case of a $10$cm glass object, such as the mirrors used in gravitational wave detectors, environment temperatures below 300mK would be required to avoid localization via blackbody scattering on the time scale relevant for entanglement generation.\\
The second scenario, gravitationally coupled oscillators, is facing similar constraints. Gas scattering and blackbody radiation remain as significant sources of decoherence. The bounds on the delocalization $\Delta x$ are now bounds on the minimal extension (or amount of squeezing) of the initial oscillator wavepacket. That this will be experimentally demanding can be seen by sticking to our previous example: achieving a wavepacket size $\Delta x>2\mu$m for a $R=75$nm silica particle as used in recent experiments ($\omega_0/2\pi\approx 10^5$Hz) would require to extend the ground-state wavepacket $\sigma\approx 3$pm by a factor of at least $6\times 10^5$. Compared to the CSIGN approach, additional requirements need to be met. Trapped oscillators will dissipate energy to the environment at a rate $\gamma$ (e.g. via photon recoil, trapped magnetic flux, etc.), which results in thermal decoherence at a rate $\Gamma_{th}=\gamma(k_BT_e/\hbar\omega_0)$. Fluctuations in trap position and frequency will add heating rates \cite{Weiss2020} $\Gamma_{x}=\pi\omega_0^2S_x(\omega_0)/(4\sigma_0^2)$ and $\Gamma_{\omega}=\pi\omega^2S_x(2\omega_0)/16$, with $S_{x,\omega}$ being the power spectral densities of position and frequency fluctuations at the relevant frequencies in the system, respectively. Current quantum experiments with massive levitated particles \cite{Delic2020,Magrini2021,Tebbenjohanns2021} are bounding these values to $\sqrt{S_x(\omega_0)}<10^{-16}$ m$\cdot$Hz$^{-0.5}$ and $\sqrt{S_{\omega}(2\omega_0)}<10^{-4}$ Hz$^{-0.5}$ in the frequency regime\footnote{In the absence of any other low-frequency noise sources, position noise $\sqrt{S_x(\omega)}$ is expected to scale with $1/\omega^2$. Extrapolation to below 100Hz will likely require a dedicated laboratory protected against other vibrational noise sources or LIGO-scale vibration isolation.} around $\omega_0/2\pi=10^5$Hz. At last the coupling needs to overcome the thermal excitations in the oscillator system. Concretely, in the absence of dissipation, the logarithmic negativity $E_N$ that detects entanglement in the system for $E_N>0$ is given by $E_N\approx (4g_s/\omega-4\bar{n})/\mathrm{ln}2$ \cite{Ludwig2010}, resulting in entanglement for $\Gamma_{\mathrm{ent}}>\bar{n}\omega_0$, or $R^3\Delta x^2\rho^2>(2/\pi)\times10^{-23}\bar{n}\omega_0$. This condition can be further relaxed for example by taking into account time-dependent interactions, which can significantly alter the system dynamics \cite{Galve2010}. 
Coming back to examples: optically trapped nanoparticles  ($R=75$nm, $\rho=2\times10^3$kg$\cdot$m$^{-3}$, $\omega_0/2\pi=1\times 10^5$Hz) in the quantum regime have been realized with a state purity $\bar{n}\approx 0.5$ ($\bar{n}$: harmonic oscillator thermal occupation) and with photon-recoil limited dissipation as low as $\gamma\approx1\times 10^{-3}$Hz. For these parameters, entanglement can be generated for $\Delta x>4$cm, when all other sources of decoherence can be neglected (which is the case for $p<10^{-6}$Pa, $T_e<40$K, $T_i<4$K, $\sqrt{S_x(\omega_0)}<10^{-15}$m$\cdot$Hz$^{-1/2}$, $\sqrt{S_{\omega}(2\omega_0)}<10^{-3}$Hz$^{-1/2}$).\footnote{Note that these numbers only define the experimental requirements once the initial conditions have been implemented. Preparing the initial states can require even more stringent environment conditions, see for example the wavepacket expansion protocol analyzed in \cite{Weiss2020}. Also note that for $\Delta x > d$ higher-order terms in the Taylor expansion of the interaction need to be taken into account.} The required large wavepacket size is a direct consequence of the energy penalty for generating entanglement between harmonic oscillators with non-unit purity. Relaxing these initial conditions to experimentally realistic values will likely require much better cooling in combination with non-stationary entanglement protocols \cite{Galve2010} or effective free-fall configurations \cite{Weiss2020}. Planck-mass lead spheres would likely be suspended at lower frequencies, say $\omega_0/2\pi=1\times 10^3$Hz. For an initial state purity of $\bar{n}\approx 0.5$ entanglement generation then requires $\Delta x>30$nm (along with $\gamma<10^{-5}$Hz, $p<10^{-15}$Pa, $T_e<6$K, $T_i<8$K, $\sqrt{S_x(\omega_0)}<10^{-23}$m$\cdot$Hz$^{-1/2}$, $\sqrt{S_{\omega}(2\omega_0)}<10^{-2}$Hz$^{-1/2}$). Even though this appears to be a significantly smaller displacement when compared to the nanosphere example, the relative wavepacket expansion $\eta=\Delta x/\sigma_0$ ($\sigma_0$: ground state size) is beyond current experimental reach in both cases. For example, the requirements $\eta>10^9$ for nanoparticles and $\eta>10^7$ for Planck-mass spheres are in stark contrast to the thus far experimentally achieved values of $\eta\approx1.2$ \cite{Wollman2015,Lecocq2015,Pirkkalainen2015}.\\
Finally, instead of using levitated objects, entanglement could also be generated between mechanically suspended oscillators. A recent experiment has demonstrated gravitational coupling between two millimetre-sized gold spheres using a torsional pendulum operating in the mHz-regime \cite{Westphal2020}. Assuming one can extend these parameters ($R\approx10^{-3}$m, $\rho\approx2\times10^4$kg$\cdot$m$^{-3}$,$\omega_0/2\pi\approx10^{-2}$Hz) to the quantum regime ($\bar{n}=0.5$) it would only require $\Delta x>900$fm to generate entanglement. However, since the respective entanglement rates are extremely small, this comes at the cost of unrealistic pressure requirements ($P<10^{-22}$Pa). Using $p<10^{-15}$Pa from above results in a consistent requirement set $\Delta x>2$nm, or a wavepacket expansion $\eta>3\times10^5$, in combination with $\gamma<3\times10^{-8}$Hz ($Q>2\times10^{6}$), $p<10^{-15}$Pa, $T_e<2.8$K, $T_i<10$K, $\sqrt{S_x(\omega_0)}<10^{-16}$m$\cdot$Hz$^{-1/2}$, $\sqrt{S_{\omega}(2\omega_0)}<10^{4}$Hz$^{-1/2}$. Similarly, if two  kilogram-scale mirrors from current gravitational wave detector experiments ($R=10^{-1}$m,$\rho=2\times10^3$kg$\cdot$m$^{-3}$, $\omega_0/2\pi\approx10^{2}$Hz) could be operated at conditions $\bar{n}<0.5$, $p<10^{-15}$Pa, $T_e<0.4$K, $T_i<5$K and $\gamma<80$Hz ($Q>10$), the generation of gravitationally induced entanglement would require a wavepacket size $\Delta x>1.2$nm, or an initial  wavepacket expansion of $\eta>10^{10}$.\\
Clearly, these considerations and related studies\cite{Krisnanda2020,Weiss2020,Pino2016,Romero-Isart2011,Rijavec2021} show that we are still far away from realistic experiments and that we need better ideas (or more patience) for going the next step.

\section{Conclusion}
The topics discussed at the Chapel Hill conference included two of the most outstanding questions in gravitational physics: \emph{Do gravitational waves exist?}, and \emph{Do we require a quantum description of gravity?}. At that time, both questions were of theoretical nature and experimental approaches did not exist beyond the power of thought experiments. In Feynman's words: ``\emph{There exists, however, one serious difficulty, and that is the lack of experiments. Furthermore, we are not going to get any experiments, so we have to take a viewpoint of how to deal with problems where no experiments are available.}" \cite{DeWitt2011}.

Today, more than 60 years later, we have an experimental answer to the first question. Laser interferometric gravitational wave detectors around the world unambiguously show almost on a daily basis that gravitational energy is radiated \cite{Abbott2016} and gravitational wave astronomy is now opening a completely new field in physics. In a parallel development, quantum experiments involving massive macroscopic objects have reached a level of maturity that we can confidently begin to tackle the second question. One possibility is to follow Feynman's initial idea and entangle objects via the gravitational field\footnote{I refer to this as the 
``level-1" excitement on the Dvali scale: the experiment shows that one cannot average space-times associated to a quantum source mass. Next levels of increasing order include the demonstration of the effect for a propagating field (level 2) \cite{Christodoulou2022} and the inclusion of field degrees of freedom, e.g. loss of entanglement visibility via entanglement with the field (level 3). This classification follows from discussions with Gia Dvali at the workshop ``Primordial black holes, de Sitter space and quantum tests of gravity" hosted by the Hamburg Academy of Sciences in February 2019.}. We have seen that the known mechanisms of decoherence impose strict and demanding limits on the experimental boundary conditions for this approach. And even though current experiments operate far away from these regimes, decoherence can in principle be tackled until we bump into a fundamental (or technological) show stopper. Thus far, this has not been the case. It is, however, also fair to say that optimistic claims about the technical feasibility of these experiments are -at least from today's view- highly exaggerated. It is certainly worthwhile to also investigate other phenomena involving coherent source mass distributions that provide us with observations that are inconsistent with what we expect from a classical (fixed) space-time metric \cite{Unruh1984}. Most likely, the outcome of such experiments will be consistent with the predictions of an effective quantum field theory. Beyond the possibility that this is not the case, the actual benefit of this undertaking lies in ruling out semiclassical models of gravity and, probably more importantly, provide empirical evidence that gravity in fact requires a quantum description. An even more thrilling perspective is to gain access to the field degrees of freedom of gravity, either directly via optical clocks \cite{Castro-Ruiz2015} or indirectly by observing decay of entanglement between the masses through additional entanglement with the field. This represents an even more demanding experimental challenge, but who knows what the next 60 years will bring.

\section{Acknowledgements}
I am grateful to my colleagues in both the quantum and the gravity community for many insightful discussions, in particular on formulating meaningful questions at the interface of these two fields. I also thank Philip Schmidt and Klemens Winkler for double-checking my numerical estimates. This work was supported by the European Research Council (ERC), Grant 649008, by the University of Vienna, the Research Platform TURIS, and by the Austrian Academy of Sciences.


%

\end{document}